\documentstyle[12pt]{article}

\begin{document}

\begin{flushright}
Preprint CAMTP/94-8\\
September 1994\\
\end{flushright}

\begin{center}
\large
{\bf Geometry of high-lying eigenfunctions in a plane billiard system having
mixed type classical dynamics}\\
\vspace{0.3in}
\normalsize
Baowen Li\footnote{e-mail Baowen.Li@UNI-MB.SI}
 and Marko Robnik\footnote{e-mail Robnik@UNI-MB.SI}\\

\vspace{0.2in}
Center for Applied Mathematics and Theoretical Physics,\\
University of Maribor, Krekova 2, SLO-62000 Maribor, Slovenia\\
\end{center}
\vspace{0.4in}
\normalsize
{\bf Abstract.}
In this work we study the geometrical properties of the  high-lying
eigenfunctions (200,000 and above) which are deep in the
semiclassical regime. The system we are analyzing is the billiard system
inside the region defined by the quadratic (complex) conformal map
$w = z + \lambda z^{2}$ of the unit disk $|z| \le 1$ as introduced
by Robnik (1983), with the shape parameter value
$\lambda = 0.15$, so that the billiard is still convex and has KAM-type
classical dynamics, where regular and irregular regions of classical
motion coexist in the classical phase space.
By inspecting 100 and by showing 36 consecutive numerically calculated
eigenfunctions we reach the following conclusions:
(1) Percival's (1973) conjectured classification in regular and irregular
states works well: the mixed type states "living" on regular {\em and}
irregular regions disappear in the semiclassical limit.
(2) The irregular (chaotic) states can be strongly localized due to the slow
classical diffusion, but become fully extended in the semiclassical limit
when the break time becomes sufficiently large with respect to the classical
diffusion time.
(3) Almost all states can be clearly associated with some relevant
classical object like invariant torus, cantorus or periodic orbits.
This paper is largely qualitative but deep in the semiclassical limit and as
such it is a prelude to our next paper which is quantitative and numerically
massive but at about ten times lower energies.
\\\\

PACS numbers: 05.45.+b, 03.65.Ge, 05.40.+j, 03.65.-w
\\\\
Submitted to {\bf Journal of Physics A}

\normalsize
\vspace{0.3in}
\newpage

\section{Introduction}

The stationary problem in quantum chaos comprises the statistical properties
of energy spectra, the statistical properties of the matrix elements of other
observables, and of geometric structure (morphology) of the eigenfunctions and
their statistics.  Some recent reviews are in Gutzwiller's book (1990), in
(Giannoni {\em et al} 1991), and in (Casati {\em et al} 1993), and also in
(Berry 1983). As for the
energy spectra we have massive numerical and experimental evidence
for the existence of universality classes of spectral fluctuations described
e.g. in (Robnik 1994), supplemented by heuristic and intuitive arguments
as well as more rigorous approaches based on applying the trace formulae
e.g. in (Berry 1985) and especially in recent works by Steiner (1994), and by
Aurich {\em et al} (1994).
In performing that kind of analysis the goodness of semiclassical
approximations still has to be carefully assessed, although Prosen and Robnik
(1993a) demonstrate that semiclassics will correctly describe the statistical
measures especially at large energy ranges even if it fails to predict
the individual energy levels.
These latter works deal with completely chaotic systems (ergodic, mixing and
having positive K-entropy) whilst the situation in generic KAM-like systems
with mixed classical dynamics is much more complicated but nevertheless
almost entirely understood in the recent works (Prosen and Robnik 1994a,b).
\\\\
The structure and the statistical properties of the eigenfunctions are not
so well understood, especially in the transition region of mixed classical
dynamics - which is the subject of the present work, whereas in the two
extreme cases of complete integrability on one side, and the complete
ergodicity on the other side, much more is known. In our recent paper
(Li and Robnik 1994a, henceforth referred to as LR) we have analyzed the
high-lying eigenfunctions in the completely ergodic regime and confirmed some
major theoretical predictions.
\\\\
In order to understand the wavefunctions especially in the semiclassical
limit it is intuitively very appealing to use the so-called {\em
Principle of Uniform Semiclassical Condensation} (PUSC) of the Wigner functions
(of the eigenstates) which is implicit in (Berry 1977a),
(Robnik 1988), and was used in LR: As $\hbar\rightarrow 0$
we assume that the Wigner function of a given eigenstate uniformly
(ergodically) condenses on the classical invariant object on which the
classical motion is ergodic and which supports the underlying quantal state.
Such an object can be e.g. an invariant torus, a chaotic  region as a
proper subset of the energy surface, or the entire energy surface if the
system has ergodic dynamics there.
\\\\
In classically integrable systems the eigenfunctions possess a lot of
ordered structure {\em globally} and {\em locally}: Applying PUSC the
average probability density in the configuration space is seen to be
determined by the projection of the corresponding quantized invariant
torus onto the configuration space, which implies the global order.
Moreover, the local structure is implied by the fact that the wavefunction
in the semiclassical limit is locally a superposition of a finite number
of plane waves (with the same wavenumber as determined by the classical
momentum).
\\\\
In the opposite extreme of a classically ergodic system PUSC predicts
that the average probability density is determined by the microcanonical
Wigner function (Shnirelman 1979, Berry 1977a, Voros 1979).
Its local structure is spanned by the superposition of
infinitely many plane waves with random phases and  equal wavenumber.
The random phases might be justified by the classical ergodicity and
this assumption, originally due to Berry (1977b), is a good starting
approximation which immediately predicts locally the Gaussian randomness
for the probability amplitude distribution. One major surprise
in this research was Heller's discovery (1984) of scars of unstable classical
periodic orbits in classically ergodic systems. The scar phenomenon
is of course a consequence of subtle correlations in the quantal phases.
This has been analyzed and discussed by Bogomolny (1988) and Berry (1989)
in the context of the Gutzwiller periodic orbit theory. The insufficiency
of the single-periodic-orbit theory of scars has been discussed by Prosen
and Robnik (1993b) in a study of the transition region between integrability
and
chaos. In the latter work P. and R. have emphasized the incompleteness of the
present day semiclassical approximations (Gutzwiller's method and torus
quantization method) in describing the individual eigenstates. Their
failure to predict the individual energy levels has been demonstrated in
(Prosen and Robnik 1993a), see also (Boasman 1994) and (Szeredi {\em et al}
1994) for discussions and related results. See also Provost and Baranger
(1993).
\\\\
In the generic case of a KAM-like system with mixed classical dynamics
the application of PUSC is again very useful and has a great qualitative
predictive power.
Here the states can be classified as either regular (they "live"
on a quantized invariant torus) or irregular (they "live" on a chaotic
invariant region), quite in agreement with Percival's (1973) speculative
prediction, which has been recently carefully re-analyzed by Prosen and
Robnik (1994a). In this case PUSC implies asymptotic ($\hbar\rightarrow 0$)
statistical independence of level series (subsequences) associated with
different regular and irregular components. This picture has been used
by Berry and Robnik (1984) to deduce the resulting energy level statistics
in such generic Hamilton systems with mixed  classical dynamics, especially
the level spacing distribution.
In the recent work Prosen and Robnik (1994a,b) have numerically confirmed
the applicability of the Berry-Robnik theory as the asymptotically
exact theory and also explained the Brody-like
behaviour (as discovered and described in (Prosen and Robnik 1993c)) before
reaching the far semiclassical limit.
\\\\
In our present paper, which is an extensive, systematic and numerically massive
work, we try to phenomenologically classify the variety of eigenstates in the
KAM-like regime of mixed classical dynamics. We do so in sections 3 and 4 by
a survey of about 100 mostly consecutive eigenfunctions of even parity
which start at about 100,000th state, of which we show and discuss a block of
36 consecutive states in configuration and in phase space in figures 2-4(a,b).
In the subsequent sections we shall
illustrate, discuss and confirm the theoretical pictures outlined above.
The main objective of the present paper is to perform clear classification of
eigenstates at the qualitative level but conceptually precise and therefore as
high as possible in the semiclassical limit, so that it can serve as a prelude
to our next paper (Li and Robnik 1994c,d) where we perform an extensive and
careful quantitative analysis of 4,000 consecutive eigenstates (no missing
states) but at ten times lower energies.

\section{The billiard system and the numerical technique}

The domain ${\cal B}_{\lambda}$ (in w-plane) of our 2-D billiard system is
defined by
the complex quadratic conformal map of the unit disk (in z-plane) onto the
complex w-plane, namely
\begin{equation}
{\cal B}_{\lambda} = \{w| w = z + \lambda z^2,\quad |z| \le 1\}.
\label{eq:cm}
\end{equation}
as introduced by Robnik (1983,1984) and further studied by Prosen and Robnik
(1993c,1994b). See also (Hayli {\em et al} 1987) and (Bruus and Stone 1994,
Stone and Bruus 1993a,b), (Frisk 1990). (Most people in the field
call this system Robnik billiard, or the Robnik model.)
As the shape parameter $\lambda$ changes from $0$ to $1/2$ this system
goes from the integrable case of the circular billiard $\lambda=0$
continuously through a KAM-like regime $0 <\lambda <1/4$
to an almost ergodic regime at  larger $\lambda$, becoming rigorously
ergodic at least at $\lambda=1/2$, where a cusp singularity appears at
the boundary point $z=-1$, mapped onto $w=-1+\lambda$, where
the mapping $w=w(z)$ is then no longer conformal, since $dw/dz=0$ there.
Because the boundary is sufficiently smooth, in fact analytic for
all $0 \leq \lambda < 1/2$, the KAM theorem
applies {\em provided} the boundary is convex. If the boundary is not convex
but
still analytic the KAM theory does {\em not} apply because the bouncing map
(Poincar\'e map) is {\em not} continuous in such a billiard.
At $0 \leq \lambda < 1/4$ the boundary is convex with positive curvature
everywhere and therefore the Lazutkin like
caustics and invariant tori (of boundary glancing orbits) exist, in
agreement with the KAM theorem (Lazutkin 1991,1981).
If the smooth boundary of a convex plane billiard has a point of zero
curvature, then it has been proven by Mather (1982,1988) that the
Lazutkin caustics and the associated invariant tori do not exist.
In our  billiard this happens at $\lambda=1/4$ for $z=-1$, i.e.
$w=-1+\lambda$.
Therefore, at $\lambda \ge 1/4$ the billiard was speculated (based on the
applicability of the Mather theorem and nonapplicability of the
KAM-theorem and on numerical
evidence in (Robnik 1983)) to become ergodic, which has been disproved
by Hayli {\em et al} (1987): Close to $\lambda\ge 1/4$ there are still some
stable periodic orbits surrounded by very tiny stability islands.
On the other hand, for $\lambda=1/2$ (the cardiod billiard, having the
above mentioned cusp singularity) the ergodicity
and mixing have been rigorously proved by Markarian (1993). See also
Wojtkowski (1986). Nevertheless,
at large values of $\lambda$, say $\lambda=0.375$ (which we studied in LR)
the numerical evidence does not exclude the possibility of rigorous ergodicity:
If there are some tiny regions of stability, then they must be so small
that they cannot be detected at large scales, as demonstrated in
(Li and Robnik 1994b), where we also show that the ergodicity may be expected
for all $\lambda\ge 0.2775$: If there is some very tiny stability
island (in the bounce map), then its relative area must be smaller
than $5.10^{-7}$.
\\\\
As in LR we want to calculate and analyze the high-lying states for our
billiard as high as 100,000th state of even parity (which means 200,000th state
when counting both parities), but this time in the transitional regime of mixed
type classical dynamics such as exemplified by our billiard at $\lambda =
0.15$. In order to achieve this goal we have to use a sophisticated and
powerful numerical technique and the best possible computer. We have used the
Convex C3860 supercomputer and the successful method to reach our goal is
our implementation of the Heller's (1991)
plane wave decomposition method described
in detail in LR. While the reader is referred to LR for all technical details
we should reassure him or her that many and all possible tests of numerical
accuracy have been performed: The energy levels are accurate within at least
$1/1000$ of the mean level spacing\footnote{The step size in a single search
(run) was $1/10$ of the mean level spacing, but we had many runs (with
different parameter values such as e.g. the interior point) and each time
an eigenvalue was captured we reached then the claimed accuracy.}
, and the eigenfunctions are accurate within
at least $1/1000$ of the average (modulus of the) probability amplitude
$1/\sqrt{{\cal A}}$, where ${\cal A} = \pi (1+2\lambda^{2})$ is the area of
the billiard.
\\\\
We have calculated more than 100 mostly consecutive even parity eigenfunctions
starting from the Weyl index 100,000. By Weyl index we mean the estimated
counting index of even parity states which as a function of energy is obtained
by applying the Weyl formula with perimeter and curvature corrections; see
equation (4) in our previous paper LR. Although our method allows us to
accurately calculate some high-lying states it does {\em not} guarantee
- unlike the diagonalization techniques - that we have
collected all the states within a given energy interval.
In fact,  typically we do
miss some states, especially pairs of almost degenerate states, so that even
after many runs and careful checks the fraction of missing levels can be as
high as 8\%. Therefore this method is certainly not suitable for performing
e.g. reliable level statistics,  but it nevertheless makes it possible to
watch and
analyze some high states deep in the semiclassical limit. In this regard it is
complementary to the conformal mapping diagonalization technique introduced by
Robnik (1984) and later used by Berry and Robnik (1986), Prosen and Robnik
(1993b,1994b) and by many others. In fact we have checked the two methods
against each other by verifying the accuracy of the energy levels as high as
10,000 where the double precision (16 digits) of the machine was established
for not too large $\lambda$. (It is our definite experience that the numerical
effort to calculate a quantum energy level increases substantially as the
degree of the classical chaos increases such as e.g. with increasing
$\lambda$.)
In another test of accuracy we were able to exactly (machine double
precision, i.e. 16 digits) reproduce the exact eigenenergies of the
analytically solvable rectangular billiard. Some of the available eigenenergies
of the Heller's stadium in the literature were likewise exactly reproduced.
Also, the wavefunctions for states as high as about 2,000, obtained by the
two different methods, have been verified to agree
within the graphical resolution. Of course we have performed many other
selfconsistent tests of accuracy of our present method e.g. by changing many
parameters  of the method, like the interior point, the boundary points and
the number of plane waves, which convinced us that the claimed accuracy (see
above) has been actually reached for the highest eigenstates.
\\\\
In spite of these difficulties we are fairly confident that we have gathered
all consecutive 36 states between the estimated Weyl index 100,008 and 100,043,
whose eigenenergies as their unique labels are given in table 1. We
believe that this is the first complete sample of consecutive high-lying
eigenfunctions in the regime of mixed type classical dynamics.
It is presented in  the configuration space and in the phase space in
figures 2(a-b), 3(a-b) and 4(a-b). Now we want to discuss this
phenomenological material, which we call "the gallery of eigenstates".

\section{The gallery of eigenstates}

The wavefunctions we are looking at are the eigenfunctions of the Schr\"odinger
equation (Helmholtz equation):
\begin{equation}
\Delta \Psi + E \Psi = 0, \qquad \Psi = 0 \quad {\rm at\quad the\quad
boundary\quad of\quad {\cal B}_{\lambda}}
\label{eq: Helm}
\end{equation}
where $E = k^{2}$ is the eigenenergy and $k$ the wavenumber. (So we are using
units such that Planck's constant $\hbar=1$ and $2m=1$, where $m$ is the mass
of the point billiard particle.)

In order to investigate the eigenstates in the quantum (Wigner) phase space we
have first to define the classical phase space and the surface of section. The
usual bounce map (Poincar\'e map) in the Birkhoff coordinates (arclength versus
tangent unit velocity vector component) is not suitable for our purpose,
because $\Psi$ vanishes on the boundary. Therefore we choose the surface of
section defined by $v= {\rm Im}(w)=0$: Our surface of section is now specified
by the
crossing point coordinate $u$ on the abscissa versus the conjugate momentum
equal to the tangential component of the velocity vector of length $k$  with
respect to the line of section $v=0$.  In figure 1(a-b) we show the geometry of
the largest chaotic component for $\lambda=0.15$ (a) and for $\lambda=0.2$ (b).
We do not show further details of the KAM scenario inside the stability islands
in order not to obscure the structure of the phase space.

The Wigner function\footnote{See e.g. Wigner (1932), Takabayasi (1954), Heller
(1976, 1977), Berry (1977a).}
(of an eigenstate $\Psi(u,v)$) defined in the full phase
space $(u,v,p_u,p_v)$ is

\begin{equation}
W({\bf q,p}) = \frac{1}{(2\pi)^{2}} \int d^{2}{\bf X}
\exp(-i{\bf p}\cdot{\bf X}) \Psi({\bf q - X}/2)\Psi({\bf q + X}/2)
\label{eq:Wigner}
\end{equation}
where we have specialized to our real $\Psi$ case, and also two degrees
of freedom and $\hbar=1$. Here ${\bf q} = (u,v)$ and ${\bf p} =(p_u,p_v)$.
In order to compare the quantum Wigner functions with the classical Poincar\'e
maps on the surface of section we define the following projection of
(\ref{eq:Wigner}) given as

\begin{equation}
\rho_{SOS}(u,p_u) = \int dp_v W(u,0,p_u,p_v),
\label{eq:proj}
\end{equation}
which nicely reduces the number of integrations by one and is equal to
\begin{equation}
\rho_{SOS} (u,p_u) = \frac{1}{2\pi} \int dx \exp(ixp_u) \Psi(u+\frac{x}{2},
0)\Psi(u-\frac{x}{2},0)
\label{eq:qsos}
\end{equation}
As is well known the Wigner function and its projections are not positive
definite and indeed one typically finds small and inconvenient but
nevertheless physical oscillations around zero which seriously obscure the main
structural features.  Therefore in order to
compare the classical and quantal phase space structure it is advisable to
smooth the Wigner function or its projections (\ref{eq:qsos}) by a normalized
Gaussian kernel with a suitably adapted dispersion. Such procedure has been
introduced and used in (Takahashi 1989, Leboeuf and Saraceno 1990,
Heller 1991, Prosen and Robnik 1993c), which is Husimi type representation
but the effective area of our Gaussian kernel will be smaller than  $2\pi$.
To be more specific, we should clearly state the size of the effective action
area $h_{eff}=2\pi/\sqrt{E}$ in all our phase space plots, in particular
(2-4b), is about $1/400$ of the entire SOS area. The area of the circle at the
half maximum of our smoothing Gaussian is about nine times smaller than
$h_{eff}$.

In discussing our gallery of eigenstates we shall use the following labelling
of individual plots in figures 2-4(a,b): (nx,i,j) identifies the plot in the
i'th row and j'th column (just the standard matrix element notation) of figure
nx, where nx denotes the number of the given figure, so it can be 2a-4b. Thus
for example (3a,3,2) is the second plot in the third row of figure 3(a).
\\\\
In figures 2(a,b) when the plots in 2b are compared with the classical plot in
figure 1(a) we immediately realize that all states should be classified as
irregular (chaotic), because their smoothed Wigner functions are concentrated
inside the classical chaotic region, except for two eigenstates (2a-b,2,1) and
(2a-b,3,2). The former of these two regular eigenstates is clearly associated
with the corresponding quantized classical invariant torus, shown in figure
6(a) together with two other regular states. This observation is based on the
comparison of the geometries by eye and could be made quantitative by
performing the torus quantization.
In our survey of more than 100 states we have seen at least 5 such similar
states on classical invariant torus with winding number close to 3.
The second regular state (2a-b,3,2) is
also "living" on a classical invariant torus but we do not show that. As for
the vast majority of chaotic states we should make the preliminary general
comment that they are typically strongly localized inside the classically
available chaotic region which is immediately obvious in the phase space plots
of figure 2(b) but not necessarily in the configuration space plots of figure
2(a).
 Structurally similar localized chaotic eigenstates are (2a,1,1), (2a,1,3) and
(2a,4,1) which is uncovered also in the phase space of figure 2(b). Another set
of similar structure are (2a,2,3), (2a,3,3) and (2a,4,3) which are
concentrated in the center of the chaotic region.
The next class of similar states are (2a,1,2) and
(2a,3,1). The chaotic state (2a,2,2) is localized at the border of stability
islands as is seen in (2b,2,2) compared with 1a. The remaining state (2a-b,4,2)
of figure 2(a-b) is strongly localized in a region where classical dynamics has
been verified to be chaotic but exhibiting very slow diffusion. This quantal
localization therefore has classical origin, because, as can be easily
verified, the quantum break time is much shorter than the classical diffusion
time. The quantum break time $t_{break}=\hbar/\Delta E$ is quite generally
defined in terms of the (locally) mean level spacing $\Delta E$ (Chirikov {\em
et al} 1981).
\\\\
In figure 3(a,b) we identify five regular states: Three of them, namely
(3a-b,1,2), (3a-b,2,1) and (3a-b,3,2), are bouncing ball type in the region
close to the horizontal diametral periodic orbit, all of them being associated
with a classical quantized invariant torus: a quasiperiodic orbit on such torus
is shown in figure 6(b) and it captures especially the structure of the state
(3a,2,1); the state (3a-b,1,3) is
a whispering gallery mode (Lazutkin 1981,1991, Keller and Rubinow 1960, Walker
1978): unfortunately its oscillatory structure having
10 nodal lines ("circles") is not graphically resolved; the state (3a-b,3,3)
is similar to (2a-b,2,1) and
thus also "lives" on the invariant torus close to period 3. We have further
three sets of similar chaotic states: (3a-b,1,1) and (3a-b,2,2); then
the two centrally localized states (3a-b,2,3) and (3a-b,4,1); and finally, the
least localized chaotic states (3a-b,3,1) and (3a-b,4,3).
Notice that as the degree of chaoticity increases from the former to the latter
in the corresponding phase space plots (3b,3,1) and (3b,4,3) we observe the
tendency towards more extended chaotic states.

We should mention that an attempt to semiclassically quantize the most regular
state (3a-b,1,3), which is a whispering gallery mode, in the Keller-Rubinow
(1960) formulation, resulted in a semiclassical energy eigenvalue which differs
from the exact one by about 5\% of the mean level spacing. This experience
is not unexpected and is conform with the demonstration by Prosen and Robnik
(1993a), where they argue that the semiclassical methods (at the level of torus
quantization or Gutzwiller theory) generally fail to
predict the individual energy levels within a vanishing fraction of the mean
level spacing even in the semiclassical limit when $\hbar\rightarrow 0$.
For related developments see Boasman (1992).
\\\\
In figure 4(a,b) we have four examples of regular states: (4a-b,2,1) which is
marginally regular (please see the classical plot in figure 1(a)),
 and (4a-b,3,3)
which in fact is an excellent example of a regular state, "living" on a thin
invariant torus around a classical periodic orbit of period 5; (4a-b,4,3)
which again is a bouncing ball type state around the horizontal diametral
periodic orbit, and (4a-b,2,3) which is an example of a "survived" whispering
gallery mode. We should stress that (4b,4,3) lives on the torus which is
close to but disjoint from the torus of the bouncing ball modes. Regarding
(4b,2,3) a similar comment applies.
In the plots (4a-b,3,2) we uncover a mixed type state which
however is close to the regular state (4a-b,3,3). All the remaining states are
chaotic but some of them are strongly localized like (4a-b,1,1), (4a-b,1,2),
(4a-b,1,3), (4a-b,2,2) and (4a-b,4,1): All of them are localized in the
region of the phase space where classical dynamics is chaotic but very
slowly diffusive. The remaining two eigenstates (4a-b,3,1) and (4a-b,4,2) are
rather extended chaotic states.

\section{Further analysis of some representative eigenstates}

Apart from the states displayed in the gallery of eigenstates we have
inspected many more (more than 100) states in configuration and in phase
space. Now we want to show and discuss some of them, belonging to various
classes. In figure 5(a) we give an example of a regular state which
semiclassically would be described as a thin quantized torus close to the
classical periodic orbit of period 4, as is evident in the phase space plot of
the same eigenstate in figure 5(b). The classical quasiperiodic orbit
associated with this state is shown in figure 6(c). In this plot we see a
signature of classical probability density explaining the structure in figure
5(a). However, having in mind that the de
Broglie wavelength is about $1/278$ of the horizontal diameter of the billiard,
one recognizes that there is a phenomenon of strong destructive interference
which leads to the gaps of strongly depressed probability density as wide as
10-15 de Broglie wavelengths. Another example of a regular state close to
the stable classical periodic orbit of period 2 is shown in figure 5(c,d):
again it is a quantum state "living" on a thin quantized torus around the
diametral periodic orbit.
\\\\
In figure 7(a) we show an interesting example of a scarred chaotic state. At
first glance it has the appearance of a regular state but in reality it is
definitely irregular but strongly localized in vicinity of the supporting
{\em unstable} classical periodic orbit of period 7. This is obvious when
figure 7(b) is compared with figure 1(a).
\\\\
In figure 7(c) we see an example of a mixed type state in the sense that in the
phase space (in figure 7(d)) it is concentrated partially both on a regular
and on a chaotic
region. Such mixed type states become more and more rare in the semiclassical
limit when $\hbar\rightarrow 0$ or equivalently when $E\rightarrow\infty$.
This observation which we have checked phenomenologically in our numerical
experiments supports the correctness of Percival's (1973) classification of
semiclassical states in regular and irregular ones.
\\\\
After our qualitative review of the variety of eigenstates we want to give some
quantitative statistical characterization. A similar analysis  of probability
amplitude distribution of strongly chaotic eigenstates (at $\lambda=0.375$)
has been published in LR, where the appropriateness of the Gaussian random
model for the probability amplitude distribution $P(\Psi)$ has been confirmed.
Similar results have been published in (Aurich and Steiner 1993).
$P(\Psi)$ is the probability density of finding an amplitude $\Psi$ inside an
infinitesimal interval $(\Psi,\Psi+d\Psi)$. Of course, in regular states or
strongly localized chaotic states there are large regions in configuration
space
where the probability density almost vanishes, and that gives rise to a delta
spike in the corresponding $P(\Psi)$ at $\Psi=0$. Therefore in the transition
region of states going from localized to extended chaotic we shall see a
gradual decrease of the central delta spike and the tendency of $P(\Psi)$
towards the Gaussian random model
\begin{equation}
P(\Psi) = \frac{1}{\sqrt{2\pi}\sigma} \exp\{-\frac{\Psi^2}{2\sigma^2}\},
\label{eq:Gaussian}
\end{equation}
where according to PUSC $\sigma^2=1/{\cal A}=1/(\pi(1+2\lambda^2))$.
In figure 8 we show such a transition for two states from the gallery of
eigenstates, namely (4a,1,3) and (2a,1,2), in figure 8(a,c) and 8(b,d),
correspondingly. The plots 8(a,b) are $P(\Psi)$ plots obtained by covering the
configuration space with about 250,000 grid points. In plots 8(c,d) we show the
corresponding cumulative distributions $I(\Psi) =\int_{-\infty}^{\Psi}dxP(x)$.
The former of these two states is clearly strongly localized chaotic whereas
the
latter is already surprisingly close to the Gaussian random model even
though it is not yet completely extended chaotic state.
\\\\
In order to illustrate the importance of the classical diffusion in the
classical chaotic regions of the phase space for the quantum localization of
the chaotic eigenstates we should remind the reader that quite generally the
quantum evolution follows the classical dynamics up to the break time, which is
by definition equal to $t_{break} = \hbar/\Delta E$ ( Chirikov {\em et al}
1981), where $\Delta E$ is the mean energy level spacing.
After the break time the quantum diffusion generally stops resulting in
a localization due to quantum interference effects.
(The reason is that in the quantum time evolution of a purely bound system
the discreteness of the spectrum of the quantum evolution operator starts
to manifest itself only at times larger than the break time.)
If the classical diffusion time is much shorter than the break time then
the quantum evolution  of an initially localized state can reach full
extendedness before the break time. We have verified that this inequality is
indeed strongly violated  at $\lambda=0.15$. Therefore, for this shape
of the billiard domain the vast majority of  chaotic states are strongly
localized as demonstrated in the gallery of eigenstates of section 3.
In order to show the approach to the extended chaotic regime we show in figure
9(a-f) three quite typical chaotic eigenstates for the billiard at
$\lambda=0.2$ all of which are fully extended chaotic as is clearly evident
when the phase space plots 9(b,d,f) are compared with figure 1(b). Of course,
we have verified that here now the break time is much larger than the classical
diffusion time, so that the extendedness is reached before the break time.

\section{Discussions and conclusions}

We believe that our present work reveals the structural richness  of
eigenfunctions in quantum systems deep in the semiclassical regime having mixed
classical dynamics. In surveying more than 100 high-lying states around and
above the 100,000th state of even parity we have persuasively demonstrated that
Percival's (1973) classification in regular and irregular states works well.
While regular states are associated with some quantized classical invariant
tori the chaotic states do not necessarily occupy the entire classically
accessible chaotic region, but can be instead strongly localized
especially in cases of slow classical diffusion where the break time is
shorter than the classical diffusion time. This qualitative observation is
demonstrated in sections 3 and 4, where we also show the approach to the
uniform extendedness in chaotic eigenstates when the above mentioned inequality
is reversed.
Thereby we have also clearly verified the validity of the principle of the
uniform semiclassical condensation outlined in the introduction.
A more quantitative study of this aspect would require to describe the
localization or extendedness at sufficiently many billiard shapes such that we
would densely cover the transition from strong localization characterized by
the inequality $t_{break}\ll t_{diff}$ to full extendedness characterized by
$t_{break}\gg t_{diff}$, where $t_{diff}$ is the classical diffusion time.
Unfortunately the present day semiclassical methods and
approximations are not yet good enough to predict (the eigenvalue and the
structure of) the individual eigenstates, c.f. (Prosen and Robnik 1993a).
Nevertheless a lot of the structure of the eigenfunctions can be associated
and qualitatively explained {\em a posteriori} by semiclassical
quantization which
thus provides the understanding of the various  classes of generic behaviour,
such as the regular and irregular states, and the further subclassification of
the latter into localized, scarred and extended states.
\\\\
We feel that the present
numerical work is a challenge to improve the semiclassical methods to the
extent that they would have the potential of predicting the individual states.
The first step in this direction has been recently undertaken by Gaspard and
Alonso (1993) where they have worked out the corrections to the leading
term embodied in the Gutzwiller trace formula.
Further numerical and theoretical work should be done to improve our knowledge
about the statistical properties of the eigenstates.  For example, one could
count the fraction of regular states within a sufficiently large block of
consecutive high-lying eigenstates, which according to the rigorous theory of
Lazutkin (1981,1991) concerning the semiclassical asymptotics should be
precisely equal to the fractional volume of the regular components in the
classical phase space.\footnote{Our gallery of 36 consecutive
eigenstates is the only block
of states for which we are fairly confident that there are no missing states.
11 of these eigenstates are regular so that their relative fraction is 30.6\%
which - surprisingly - is not too far from the classical fractional volume of
the regular component in the classical phase space which according to Prosen
and Robnik (1993c) is equal to 36\%.}
This fact is e.g. one major assumption in the
Berry-Robnik (1984) theory. For this to end we need to have a better numerical
method with no missing of eigenstates ensuring a significant statistics, which
at present we do not (yet) possess for such high-lying states, whereas for low
states (say, up to 10,000) we certainly could use our conformal
diagonalization technique (see e.g. Prosen and Robnik (1993c)) but then we are
probably not sufficiently far in the semiclassical limit for an unambigous
classification in regular and irregular states. This is studied and discussed
in detail in our next paper (Li and Robnik 1994c,d), where we successfully
separate regular and irregular energy levels, using the dynamical criterion of
comparing the classical and quantal phase space plots on SOS, and investigate
the level statistics of the two level sequences separately. But there is still
much more work to be done especially such as a more quantitative analysis on
points (1-3) raised and listed in the abstract.

\section*{Acknowledgments}

We thank Toma\v z Prosen for a few computer programs and assistance in
using them. One of us (MR) acknowledges stimulating discussions with
Oriol Bohigas, Boris V. Chirikov, Marcos Saraceno and Hans A. Weidenm\"uller.
Finally we thank the second of the two referees for the
careful, constructive and critical report.
The financial support by the Ministry of Science and
Technology of the Republic of Slovenia is gratefully acknowledged.

\vfill
\newpage
\section*{References}
Aurich R and Steiner F 1993 {\em Physica D} {\bf 64}, 185\\\\
Aurich R, Bolte J and Steiner F 1994 {\em Phys. Rev. Lett.} {\bf 73}, 1356\\\\
Berry M V 1977a {\em Phil. Trans. Roy. Soc. London} {\bf 287} 237\\\\
Berry M V 1977b {\em J. Phys. A: Math. Gen.} {\bf 10} 2083\\\\
Berry M V 1983 in  in {\em Chaotic Behaviour of Deterministic Systems
(Proc. NATO ASI Les Houches Summer School)} eds
Iooss G, Helleman R H G and Stora R (Amsterdam: Elsevier) p171\\\\
Berry M V 1985 {\em Proc. Roy. Soc. London} {\bf A400} 229\\\\
Berry M V 1989 {\em Proc. Roy. Soc. London} {\bf A423} 219\\\\
Berry M V and Robnik M 1984 {\em J. Phys. A: Math. Gen.} {\bf 17} 2413\\\\
Berry M V and Robnik M 1986 {\em J. Phys. A: Math. Gen.} {\bf 19} 649\\\\
Boasman P A 1994 {\em Nonlinearity} {\bf 7} 485\\\\
Bohigas O, 1991  in {\em Chaos and Quantum Systems (Proc. NATO ASI Les Houches
Summer School)} eds M-J Giannoni, A Voros and J Zinn-Justin,
(Amsterdam: Elsevier) p87\\\\
Bogomolny E B 1988 {\em Physica D} {\bf 31} 169\\\\
Bruus H and Stone A D 1994 {\em Phys. Rev. B} {\bf 50} 18275\\\\
Casati G, Guarneri I and Smilansky U eds. 1993
{\em Quantum Chaos} (North-Holland) \\\\
Chirikov B V, Shepelyansky D L and Izrailev F M  1981 {\em Sov. Sci. Rev.}
{\bf C2} 209\\\\
Frisk H 1990 Nordita {\em Preprint}.\\\\
Gaspard P and Alonso D 1993 {\em Phy. Rev. E} {\bf 47} R3468\\\\
Giannoni M-J, Voros J and Zinn-Justin eds. 1991 {\em Chaos and Quantum Systems}
(North-Holland)\\\\
Gutzwiller M C 1990 {\em Chaos in Classical and Quantum Mechanics} (New York:
Springer)\\\\
Hayli A, Dumont T, Moulin-Ollagier J and Strelcyn J M 1987 {\em J. Phys. A:
Math. Gen} {\bf 20} 3237\\\\
Heller E J 1976 {\em J. Chem. Phys.} {\bf 65} 1289\\\\
Heller E J 1977 {\em J. Chem. Phys.} {\bf 67} 3339\\\\
Heller E J 1984 {\em Phys. Rev. Lett} {\bf 53} 1515\\\\
Heller E J 1991  in {\em Chaos and Quantum Systems (Proc. NATO ASI Les Houches
Summer School)} eds M-J Giannoni, A Voros and J Zinn-Justin,
(Amsterdam: Elsevier) p547\\\\
Keller J B and Rubinow S I {\em Ann. Phys. N.Y.} {\bf 10} 303\\\\
Lazutkin V F 1981 {\em The Convex Billiard and the Eigenfunctions of
the Laplace Operator} (Leningrad: University Press) (in Russian)\\\\
Lazutkin V F 1991 {\em KAM Theory and Semiclassical Approximations to
Eigenfunctions} (Heidelberg: Springer Verlag)\\\\
Leboeuf P and Saraceno M 1990 {\em J. Phys. A: Math. Gen.} {\bf 23} 1745\\\\
Li Baowen and Robnik M 1994a {\em J. Phys. A: Math. Gen.} {\bf 27} 5509\\\\
Li Baowen and Robnik M 1994b to be submitted to {\em J. Phys. A: Math.
Gen.}\\\\
Li Baowen and Robnik M 1994c {\em Preprint CAMTP/94-10} submitted to {\em J.
Phys. A: Math. Gen.} in December 1994\\\\
Li Baowen and Robnik M 1994d {\em Preprint CAMTP/94-11} to be submitted\\\\
Markarian R 1993 {\em Nonlinearity} {\bf 6} 819\\\\
Mather J N 1982 {\em Ergodic Theory and Dynamical Systems} {\bf 2} 3\\\\
Mather J N 1988 {\em Ergodic Theory and Dynamical Systems} {\bf 8} 199\\\\
Percival I C 1973 {\em J. Phys. B: At. Mol. Phys.} {\bf 6} L229 \\\\
Prosen T and Robnik M 1993a {\em J. Phys. A: Math. Gen.} {\bf 26} L37\\\\
Prosen T and Robnik M 1993b {\em J. Phys. A: Math. Gen.} {\bf 26} 5365\\\\
Prosen T and Robnik M 1993c {\em J. Phys. A: Math. Gen.} {\bf 26} 2371\\\\
Prosen T and Robnik M 1994a {\em J. Phys. A: Math. Gen.} {\bf 27} L459\\\\
Prosen T and Robnik M 1994b  {\em J. Phys. A: Math. Gen.} {\bf 27} 8059\\\\
Provost D and Baranger M 1993 {\em Phys. Rev. Lett.} {\bf 71} 662\\\\
Robnik M 1983 {\em J. Phys. A: Math. Gen.} {\bf 16} 3971\\\\
Robnik M 1984 {\em J. Phys. A: Math. Gen.} {\bf 17} 1049\\\\
Robnik M 1988 in {\em "Atomic Spectra and Collisions in External Fields"},
eds. K T Taylor, M H Nayfeh and C W Clark, (New York: Plenum) pp265-274\\\\
Robnik M 1994 {\em J. Phys. Soc. Japan Suppl.} {\bf 63} 131\\\\
Shnirelman A L 1979 {\em Uspekhi Matem. Nauk} {\bf 29} 181\\\\
Steiner F 1994 {\em Quantum Chaos} In {\em Schlaglichter der Forschung. Zum 75
Jahrestag der Universit\"at Hamburg 1994}, (Ed. R. Ansorge) Festschrift
published on the occasion of the 75th anniversary of the University of Hamburg,
543 (Dietrich Reimer Verlag, Berlin and Hambrug 1994).\\\\
Stone A D and Bruus H 1993a {\em Physica B} {\bf 189} 43\\\\
Stone A D and Bruus H 1993b {\em Surface Science} {\bf 305} 490\\\\
Szeredi T, Lefebvre J H and Goodings D A 1994 {\em Nonlinearity} {\bf 7}
1463\\\\
Takabayasi T 1954 {\em Prog. Theor. Phys. (Kyoto)} {\bf 11} 341\\\\
Takahashi K 1989 {\em Prog. Theor. Phys. Suppl. (Kyoto)} {\bf 98} 109\\\\
Voros A 1979 {\em Lecture Notes in Physics} {\bf 93} 326\\\\
Walker J 1978 {\em Sci. Am.} {\bf 239} 146\\\\
Wojtkowski M 1986 {\em Commun. Math. Phys.} {\bf 105} 391\\\\

\vfill
\newpage
\section*{Table}

\bigskip
\bigskip

 {\bf Table 1.} The eigenenergies of the corresponding eigenstates
 in figures 2(a-b),3(a-b),
 4(a-b). The $E=k^{2}$ are in the left-right, top-down order.\vspace{15mm}\\

\begin{tabular}{|lll|} \hline
$ 766,265.500 $ & $ 766,272.567 $ & $ 766,277.395$\\
$ 766,282.954 $ & $ 766,293.827 $ & $ 766,297.762$\\
$ 766,305.324 $ & $ 766,310.756 $ & $ 766,313.551$\\
$ 766,344.590 $ & $ 766,347.851 $ & $ 766,353.425$\\ \hline
$ 766,356.683 $ & $ 766,371.767 $ & $ 766,379.112$\\
$ 766,380.778 $ & $ 766,386.997 $ & $ 766,395.422$\\
$ 766,403.315 $ & $ 766,413.436 $ & $ 766,428.555$\\
$ 766,433.225 $ & $ 766,440.910 $ & $ 766,442.929$\\ \hline
$ 766,460.422 $ & $ 766,460.745 $ & $ 766,466.389$\\
$ 766,470.449 $ & $ 766,476.940 $ & $ 766,488.165$\\
$ 766,497.289 $ & $ 766,502.487 $ & $ 766,503.689$\\
$ 766,511.361 $ & $ 766,514.810 $ & $ 766,527.390$ \\ \hline
\end{tabular}\vspace{10mm}\\

\newpage
\section*{Figure captions}

\bigskip
\bigskip

\noindent {\bf Figure 1:}
The classical SOS of the billiard system. (a) for $\lambda=0.15$, (b) for
$ \lambda = 0.2$. For the definitions see text.

\bigskip
\bigskip

\noindent {\bf Figure 2:}
The first 12 states from the gallery of eigenstates. The eigenfunctions in
configuration space (a) and the corresponding smoothed Wigner function
$\rho_{SOS}$ (b). In (a) the contours are plotted at 8 equally
spaced steps between zero and the maximum value. In (b) the contours
start from 1/8 with the step size 1/8. The abscissa in (b) is
just the coordinate on the line of section whilst the ordinate
goes from $-\sqrt(E)$ to $\sqrt(E)$, where $E$ is the eigenenergy of the given
state. Notice please that  the classically allowed value of $p_u$ is within the
interval $[-\sqrt{E}, \sqrt{E}]$.
For labelling of individual plots please see text.

\bigskip
\bigskip

\noindent {\bf Figure 3:}
The same as figure 2 but for the second 12 states from the gallery of
eigenstates.

\bigskip
\bigskip

\noindent {\bf Figure 4:}
The same as figure 2 but for the last 12 states from the gallery of
eigenstates.

\bigskip
\bigskip

\noindent {\bf Figure 5:}
Two examples of regular states in configuration space (a,c) and in phase
space (b,d). The energy of the top one is 766,536.529 and the bottom one
is 766,569.685.

\bigskip
\bigskip
\noindent {\bf Figure 6:}
A quasiperiodic classical orbit on the invariant torus supporting a regular
eigenstate: (2a-b,2,1) in (a); (3a-b,2,1) in (b) and figure 5(a,b) in (c).

\bigskip
\bigskip

\noindent {\bf Figure 7:}
Two examples of eigenstates in configuration space (a,c) and in phase space
(b,d). The energy of the top one is 766,718.836 and the bottom one is
766,940.785. (a,b) is a scarred chaotic state whereas (c,d) is a mixed type
state.

\bigskip
\bigskip

\noindent {\bf Figure 8:}
The statistics of two eigenstates (4a-b,1,3) in (a,c) and  (2a-b,1,2) in (b,d).
In (a,b) we plot the amplitude probability density $P(\Psi)$ and in (c,d)  the
cumulative distribution $I(\Psi)$. The numerical data are shown in full line
and the theoretical Gaussian in dashed line.

\bigskip
\bigskip

\noindent {\bf Figure 9:}
Three examples of extended chaotic eigenstates for $\lambda=0.2$ in
configuration space (a,c,e) and in phase space (b,d,f). The energies are:
741,511.898 in (a,b); 741,525.937 in (c,d); 741,549.855 in (e,f). The Weyl
index of these states is: 100,017, 100,019 and 100,022, respectively.

\end{document}